# Finite Size Effects in Highly Scaled Ruthenium Interconnects

Shibesh Dutta, Kristof Moors, Michiel Vandemaele, and Christoph Adelmann

*Abstract*— Ru has been considered a candidate to replace Cu-based interconnects in VLSI circuits. Here, a methodology is proposed to predict the resistivity of (Ru) interconnects. First, the dependence of the Ru thin film resistivity on the film thickness is modeled by the semiclassical Mayadas-Shatzkes (MS) approach. The fitting parameters thus obtained are then used as input in a modified MS model for nanowires to calculate wire resistivities. Predicted experimental resistivities agreed within about 10%. The results further indicate that grain boundary scattering was the dominant scattering mechanism in scaled Ru interconnects.

## I. Introduction

FINITE size effects denote the phenomenon of the resistivity increase of metallic nanostructures when their characteristic dimensions are reduced to the order of the mean free path, and are due to the enhanced scattering of charge carriers at the interfaces and the grain boundaries [1–3]. The increased resistivity has important technological implications. Currently, Cu wires are used as interconnects in VLSI circuits and the increased resistivity due to scaling has led to problems of higher latency, increased power consumption, higher noise, and degraded reliability [4–6]. Moreover, scaling limitations of the diffusion barriers used in dual-damascene Cu interconnects reduce the volume fraction occupied by Cu in the interconnects increasing the line resistance even further [7].

Pt-group metals, especially Ru, have emerged as promising candidates to replace Cu in future interconnects [8] due to their weaker thickness dependence of the resistivity [9, 10]. Ru interconnects have demonstrated robust reliability [11–13] and potential for barrierless integration [13–15]. Although Ru wires with cross-sectional areas as small as 33 nm$^2$ have been demonstrated [16], the scattering mechanisms in such wires and their impact on the resistivity have not been studied. Moreover, existing models to predict the resistivity of metallic interconnects typically require many fitting parameters that are often non-trivial to determine experimentally [17–19].

Shibesh Dutta is with imec, 3001 Leuven, Belgium and also with the Department of Physics and Astronomy, KU Leuven, 3001 Leuven, Belgium (e-mail: shibeshdutta@gmail.com).

Kristof Moors is with the University of Luxembourg, Physics and Materials Science Research Unit, 1511 Luxembourg, Luxembourg.

Michiel Vandemaele is with the Department of Electrical Engineering, KU Leuven, 3001 Leuven, Belgium and also with imec, 3001 Leuven, Belgium.

Christoph Adelmann is with imec, 3001 Leuven, Belgium.

In this work, we describe a model for the resistivity of Ru wires that is derived from the established semiclassical model for thin films by Mayadas and Shatzkes [3]. We demonstrate the capability of this model to predict quantitatively the resistivity of Ru wires with cross-sectional areas between 30 and 130 nm$^2$ using experimentally determined linear grain boundary intercept lengths [20] in combination with the surface specularity parameter $p$ and the grain boundary reflection coefficient $R$ (that describe the nature of surface and grain boundary scattering respectively) obtained from thin film experiments. The model further allows to assess the impact of the surface and the grain boundary scattering and examine the effect of aspect ratio (AR) on the resistivity of Ru wires.

## II. Thin Film Characterization and Resistivity Modeling

First, we determine the surface specularity parameter $p$ and the grain boundary reflection coefficient $R$ for Ru films (5–30 nm) deposited by atomic layer deposition (ALD) from (ethylbenzyl) (1-ethyl-1,4-cyclohexadienyl) Ru$^{(0)}$ (EBECHRu) and O$_2$ on SiO$_2$/Si (100) both with and without a 0.3 nm thick TiN adhesion layer [21]. Further information on process chemistry and growth mechanisms can be found elsewhere [21]. All films were annealed at 420°C for 20 min in forming gas. Plan-view TEM images along with the deduced linear grain boundary intercept lengths [20] are shown in Fig. 1. Smaller linear intercept lengths (smaller grains) were observed for Ru films on 0.3 nm TiN/SiO$_2$ than on SiO$_2$ only. This may be attributed to Ti segregation into Ru films [21], which possibly occupies grain boundary interstices and increases the activation energy for grain boundary migration [22].

The thickness dependence of the resistivity of ALD Ru films on SiO$_2$ with and without the 0.3 nm TiN adhesion layer is shown in Fig. 1d. Ru films on SiO$_2$ show comparatively lower resistivity owing to their larger linear intercept lengths (larger grains). To gain further insight into the scattering mechanisms, the thickness dependence of resistivity of the ALD Ru films was modeled using the semiclassical MS model [3, 10]. Experimental linear intercept lengths were used together with the bulk electron mean free path obtained from *ab initio* calculations [9, 10] and experimental values of the bulk resistivity [23]. The bulk resistivity for Ru is anisotropic and thus the in-plane resistivity of 7.6 μΩcm was used to model Ru films on SiO$_2$ as the films were found to exhibit strong (001) texture [21]. By contrast, Ru films on 0.3 nm TiN showed no clear texture and thus a bulk resistivity of 7.1

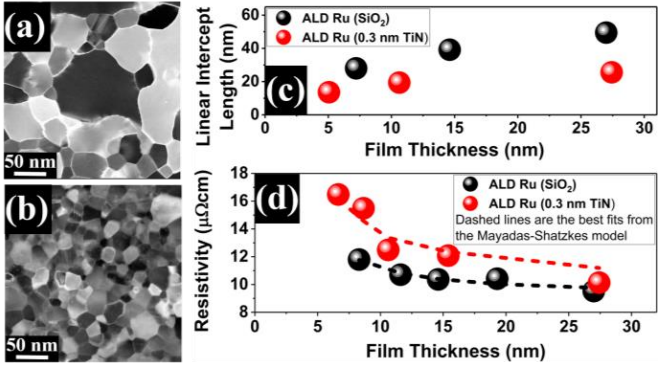

Figure 1. In-plane TEM images of ~27 nm thick ALD Ru films (a) on SiO$_2$, and (b) on 0.3 nm TiN-SiO$_2$. (c) Comparison of linear intercept lengths (d) Thickness dependence of resistivity of ALD Ru, along with the best fits from the MS model.

µΩcm was used, corresponding to a random polycrystal [21].

The parameters $p$ ($0 \leq p \leq 1$) and $R$ ($0 \leq R \leq 1$) in the MS model are phenomenological fitting parameters. $p$ is the degree of specular scattering of the charge carriers at the interfaces, with $p=1$ and $p=0$ corresponding to specular and diffuse scattering, respectively. $R$ is the reflection coefficient of the charge carriers at the grain boundaries. The model described well the thickness dependence of Ru film resistivity on both substrates. The resulting parameters are listed in Table 1. For ALD Ru films on SiO$_2$, surface scattering was found to be nearly specular ($p \sim 1$) and thus grain boundary scattering ($R = 0.40$) was the dominant contributor to resistivity. This is very similar to a report on PVD Ru films on SiO$_2$ [10] and suggests that deposition process itself does not strongly affect the surface scattering behavior. By contrast, surface scattering in Ru films on 0.3 nm TiN adhesion layers was diffuse, with $p \sim 0$. This is again similar to PVD Ru [10] where the insertion of interfacial and surface TaN layers led to diffuse scattering at the Ru interfaces, which indicates the highly sensitive nature of surface scattering and shows that even very thin TiN layers can have a strong influence. We note that due to the segregation of Ti(O,N)$_x$ during the ALD, Ti (on the order of a monolayer) is also present on the surface after deposition [21] and thus can be expected to modify the scattering at the top as well as bottom interface (i.e. in an effective Ti(O,N)$_x$/Ru/TiN stack). On TiN, the grain boundary reflection coefficient was marginally higher ($R = 0.45$) than for Ru films on SiO$_2$. The determined $R$ values obtained for ALD Ru agree well with a previous report on PVD Ru [10] and with a recent theoretical model for $R$ [24] that predicts values of ~0.5.

TABLE 1. Best fitting parameters, $p$ and $R$, along with mean free path $\lambda_0$ and bulk resistivity $\rho_0$ (Ref. [10]) used as input parameters in the MS model. The sum of squared errors (SSE) is also given for each data set. For comparison, fitting parameters for PVD Ru on SiO$_2$ and PVD Cu on TaN published elsewhere [10] are also provided.

|  | $p$ | R | $\lambda_0$ (nm) | $\rho_0$ (µΩcm) | SSE (µΩ$^2$cm$^2$) |
|---|---|---|---|---|---|
| ALD Ru on SiO$_2$ | 0.99 | 0.40 | 6.6 | 7.6 | 0.2 |
| ALD Ru on TiN | 0.01 | 0.45 | 6.6 | 7.1 | 2.9 |
| PVD Ru on SiO$_2$ [10] | 0.98 | 0.43 | 6.6 | 7.6 | 8.9 |
| PVD Cu on TaN [10] | 0.05 | 0.22 | 40.6 | 1.7 | 0.6 |

## III. RESISTIVITY MODELING OF RUTHENIUM WIRES

The MS model was developed for thin films but can be modified for rectangular wires by imposing additional set of boundary conditions at the sidewalls of the wire, i.e. by considering the wires as two "intersecting thin films" with thickness $h$ and width $w$, respectively. In the thin film MS model, the resistivity $\rho_{tf}$, can be written as $\rho_{tf} \equiv (1/\rho_{GB} - 1/\rho_{SS,GB})^{-1}$ with the first term describing grain boundary scattering and the second, surface scattering renormalized by grain boundary scattering [3, 10]. Imposing additional boundary conditions due to the finite width and treating surface scattering at the two sets of interfaces as independent, the resistivity of a wire becomes $\rho_{nw} \equiv [1/\rho_{GB} - (1/\rho_{SS,GB-h} + 1/\rho_{SS,GB-w})]^{-1}$. Using the expressions of Ref. [3], the resistivity of rectangular wires can then be written as:

$$\rho_{nw} = \left[\frac{1}{\rho_{GB}} - \left\{\frac{6(1-p)}{\pi\rho_0}\left(\left(\frac{1}{h}\int_0^{\pi/2}d\varphi\int_1^{\infty}dt\,\frac{\cos^2\varphi}{H^2}\left(\frac{1}{t^3}-\frac{1}{t^5}\right)\frac{1-e^{-k_1 tH}}{1-pe^{-k_1 tH}}\right)\right.\right.$$
$$\left.\left.+\left(\frac{1}{w}\int_0^{\pi/2}d\varphi\int_1^{\infty}dt\,\frac{\cos^2\varphi}{H^2}\left(\frac{1}{t^3}-\frac{1}{t^5}\right)\frac{1-e^{-k_2 tH}}{1-pe^{-k_2 tH}}\right)\right)\right\}\right]^{-1} \quad (1)$$

with

$$H = 1 + \frac{\alpha}{\cos\phi\sqrt{1-\frac{1}{t^2}}} \qquad \alpha = \frac{\lambda_0}{g}\frac{2R}{1-R} \qquad k_1 = \frac{h}{\lambda_0} \qquad k_2 = \frac{w}{\lambda_0}$$

To calculate the resistivity of Ru wires, the input parameters in (1) (i.e. $p$, $R$, $\rho_0$, and $\lambda_0$) were taken from Table 1 for ALD Ru on 0.3 nm TiN, the same metallization as in the Ru interconnects [16]. The linear intercept length $g = 9.3 \pm 1.5$ nm was obtained from cross-sectional TEM images along the length of the wire (Fig. 2a). Since the wires were obtained by etching perpendicular to the transport direction, it can be assumed that the intercept length is independent of the wire area. The cross-sectional area was determined by temperature-dependent resistance measurements [25]. The AR of the wires was determined from TEM images [16] for the smallest and the largest cross-sectional areas and interpolated in between. Individual height and width of the wires were then calculated from measured areas and ARs. Details of fabrication, and the results of physical and electrical characterization of the Ru wires have been published elsewhere [16].

The computed resistivities using Eq. (1) with $R = 0.45$ and $p = 0$ were about 10–15% lower than the experimental values (Fig. 2b). This may be because the Ru interconnects were only annealed after full dielectric passivation of the films. This typically leads to reduced recrystallization and smaller grains than annealing of structures with free surfaces. It has been reported for Ru films that annealing reduces $R$ in parallel to the recrystallization [10]. Thus, the observed deviation may be attributed to a slight underestimation of the $R$ value obtained from annealed ALD Ru films (with a free surface). A better agreement between model and data can be obtained for $R = 0.49$ (Fig. 2b). The model now agrees well with the experimental data for areas above 70 nm$^2$ but it still slightly underestimates the experiment for smaller areas, albeit by < 10%. Such deviations are likely due to effects than are not captured by the semiclassical MS model. In the simple MS model, the maximum contribution of surface scattering is obtained for *p* = 0, *i.e.* for fully diffuse scattering at a completely flat boundary surface. In very narrow wires, such boundary roughness effects are expected to become more pronounced due to a non-negligible probability of having sequential scattering events (*i.e.* the surface acting like a trap) or an effective reduction of the cross-sectional area. This may lead to a resistivity contribution of surface scattering even higher than for *p* = 0, which may in turn explain the systematic underestimation of the resistivity at small areas. Moreover, quantum effects (quantization of the conduction band) may also lead to an increase of the resistivity over the bulk resistivity even in absence of surface and grain boundary scattering, as shown for single crystal nanowires by *ab initio* calculations [26, 27].

Despite all these issues, reasonable agreement between experimental and modeled resistivity was already obtained using parameters obtained from thin film experiments. This indicates that the parameters obtained by Ru thin film resistivity modeling also apply (in good approximation) to Ru wires and that much simpler thin film experiments may be used to predict the resistivity of scaled interconnects with reasonable accuracy. A calculation for the case of fully specular interface scattering (*p* = 1; all other input parameters remain the same) is also shown as a dashed line in Fig. 2b. The difference between experimental data and the dashed line can be considered as the contribution of surface scattering to the resistivity. Thus, grain boundary scattering is the dominant contribution to the resistivity, except for the smallest areas (< 50 nm$^2$) where surface scattering starts to contribute significantly.

For comparison, the resistivity of Cu wires was computed as a function of the cross-sectional area using *p* = 0.05 and *R* = 0.22 as obtained from the resistivity modeling of thin Cu films reported elsewhere [10]. Other input parameters are listed in Table 1. ARs and linear intercept length were the same as that of Ru wires. The area dependence of the resistivity of Cu wires was found to be stronger than that of Ru wires due to its larger mean free path [10]. Hence, the simulations predict a crossover of the resistivities of Cu and Ru wires around a cross-sectional area of 45 nm$^2$ (for the parameters used here) below which Ru wires would have lower resistivity.

Using the semiclassical model, the effect of the AR on the resistivity of Ru wires can also be examined. The resistivity of wires at a fixed cross-sectional area (Fig. 2c) shows a rather weak dependence on the AR of the structures. Even for the smallest simulated area of 25 nm$^2$, the simulations predict only an ~8% variation in resistivity between ARs of 4 to 1 and a decrease of < 2% when the AR is reduced from 2 (a realistic value for interconnects) to 1. For the largest simulated area of 200 nm$^2$, the change in resistivity is even smaller; ~3% between ARs of 4 and 1, and < 1% when reducing the AR from 2 to 1. The weak AR dependence is due to the fact that both the height and the width of the wires are of the order of (or below) the mean free path and the surface scattering contributions due to changing height and width mostly compensate for each other. Note that the contribution of grain boundary scattering to resistivity remains constant as the linear intercept length is assumed to be independent of the AR. This suggests that the shape of the wire for a given cross-sectional area might not strongly affect the resistivity and thus a simplistic rectangular profile for the wires assumed here is a reasonable approximation.

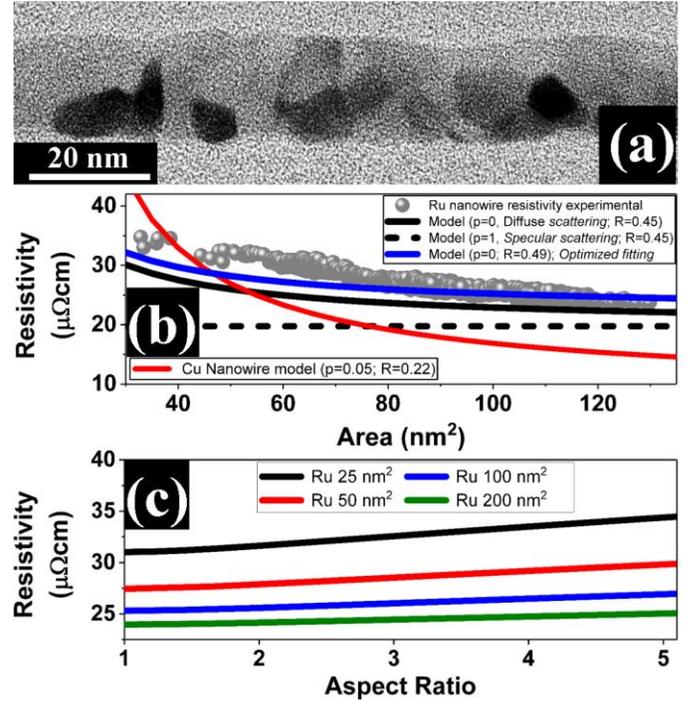

Figure 2. (a) TEM image of Ru wire (b) Calculated area dependence of resistivity of Ru wires using Ru thin film modeling parameters (*p*=0, *R*=0.45), and compared to experimental Ru wire resistivity from [16]. An optimized fit (*p*=0, *R*=0.49) is also included. The calculated resistivity in case of specular surface scattering is shown. For comparison, the calculated resistivity of Cu wires is also shown using thin film parameters (*p*=0.05, *R*=0.22) [10]. (c) Effect of the aspect ratio of wires on the resistivity for various fixed cross-sectional areas.

## IV. CONCLUSION

The resistivity of Ru wires was computed using the parameters obtained from the resistivity modeling of ALD Ru films as inputs to a modified MS model adapted for rectangular nanowires, and showed good agreement with the experimental data [20]. The proposed methodology holds the potential to predict the resistivity of any metallic interconnect structure based on modeling parameters obtained by comparably simple thin film analysis. Grain boundary scattering was the dominant mechanism contributing to the resistivity of Ru wires, except at very small cross-sectional areas where the contribution of surface scattering was significant. In addition, the aspect ratios of wires have little impact on the resistivity and the resistivity is mostly determined by cross sectional area rather than by width and height individually.

## V. ACKNOWLEDGMENT

The authors acknowledge the support by imec's industrial affiliate program on Nano-interconnects and by the National Research Fund Luxembourg (ATTRACT Grant No. 7556175).